\documentstyle[12pt]{l-aa}
\input{psfig}
\begin{document}

\thesaurus{(08.13.1; 08.14.1; 13.07.2)}
\title{Pulse Phase Spectroscopy of A~0535+26 during its 1994 
giant outburst observed with OSSE}
\author{Michael Maisack         \inst{1}
\and    J. Eric Grove           \inst{2}
\and    Eckhard Kendziorra      \inst{1}
\and    Peter Kretschmar        \inst{1}
\and    R\"udiger Staubert      \inst{1}
\and    Mark S. Strickman       \inst{2}
}
\offprints{M. Maisack, maisack@astro.uni-tuebingen.de}
\institute{Institut f\"ur Astronomie und Astrophysik , Abteilung Astronomie, 
           Universit\"at T\"ubingen, 72076 T\"ubingen, Germany
\and       E.O. Hulburt Center for Space Research, Naval Research Laboratory, 
           Washington, DC 20375, USA}
\date{Received 1997; accepted XXX}
\maketitle
\markboth{M. Maisack et al.: Pulse Phase Spectroscopy of A~0535+26}{}

\begin{abstract}

We present pulse phase spectroscopy of A~0535+26 in the energy range 
35-200~keV from OSSE observations of 
its giant outburst in 1994. We discuss the phase dependence of the 
continuum parameters and the cyclotron resonance feature (CRF) at 110 keV 
already found in the phase averaged spectrum. We find that a CRF is required at 
every phase. 
The behaviour of the line parameters with phase 
and the pulse shape indicate that the emission occurs in a pencil-beam 
geometry.

\keywords{Stars: magnetic fields; Stars: neutron; Gamma rays: observations}
\end{abstract}

\section{Introduction}

In Be X-ray binaries (BeXRB) with a neutron star as the compact object, 
mass transfer is mediated via a circumstellar disk around the companion. 
The companions are O or B stars with high rotational velocities which are 
responsible for the loss of material in the equatorial plane. 
The material in this disk, which may be present or absent for extended 
periods, flows out rather slowly (velocities of several tens of km/s) 
compared to the fast B star wind outside 
the equatorial plane. In fact, the dynamics of the material in this 
circumstellar disk may be dominated by Keplerian motion (e.g. Hanuschik 
1996).

BeXRB undergo dramatic increases in X-ray flux when the neutron star 
passes through the dense regions of the circumstellar disk near periastron 
passage and enhanced mass accretion sets in. Outbursts with luminosities 
in excess of $10^{38}$ ergs/s can be observed in these systems at these 
times, while the X-ray luminosity during quiescence is 10$^{33}$ ergs/s and 
below.

The brightest object of this class is the system A~0535+26 (HDE~245770), 
which has an orbital period of $\approx$ 110 days, and a pulsation period 
of $\approx$ 103 s. For a comprehensive review of the system properties, 
see Giovannelli and Graziati (1992). 
The system shows relatively regular outbursts occuring 
near periastron passage, and occasional giant outbursts during which the 
system becomes several times brighter than the Crab between $\approx$ 5 and 
100~keV. The system has also been found in 
prolonged off states (Motch et al. 1991).

Four giant outbursts have been observed so far; during the last 
two of these in 1989 and 1994, cyclotron lines (also called cyclotron 
resonance features , CRF) have been 
observed (Kendziorra et al. 1994, Grove et al. 1995). 
CRF in the X-ray regime are produced in very high 
magnetic fields (B $>$ 10$^{11-12}$ G), 
where the cyclotron energy becomes comparable to the electron rest mass.
The electrons can then only gyrate on discrete Landau levels. Transitions 
between these levels result in line features in the X-ray spectra.

Kendziorra et al. (1994) found two features in the phase resolved TTM/HEXE 
spectra of A~0535+26 at $\approx$ 50 and 100 keV, whereas Grove et al. (1995) 
found 
a feature only at $\approx$ 110~keV in the phase averaged spectrum during an 
outburst observed by OSSE. However, the presence of a feature at 
$\approx$~50~keV could not be ruled out due to the higher energy threshold 
of OSSE compared to TTM/HEXE. Regardless of the existence of the 
50 keV 
feature, the magnetic field of the neutron star in A~0535+26 is the highest 
that has been measured directly so far (for a review of 
cyclotron line sources, see Makishima and Mihara 1992).

In this paper, we present pulse phase resolved spectroscopy of the 1994 
outburst observed by OSSE. We discuss the variability of the continuum and 
the 110 keV line feature across the pulse, the possibility of the presence 
of a feature at $\approx$ 50 keV and physical implications for the emission 
pattern at the magnetic poles.

\begin{figure}[t]
  \psfig{figure=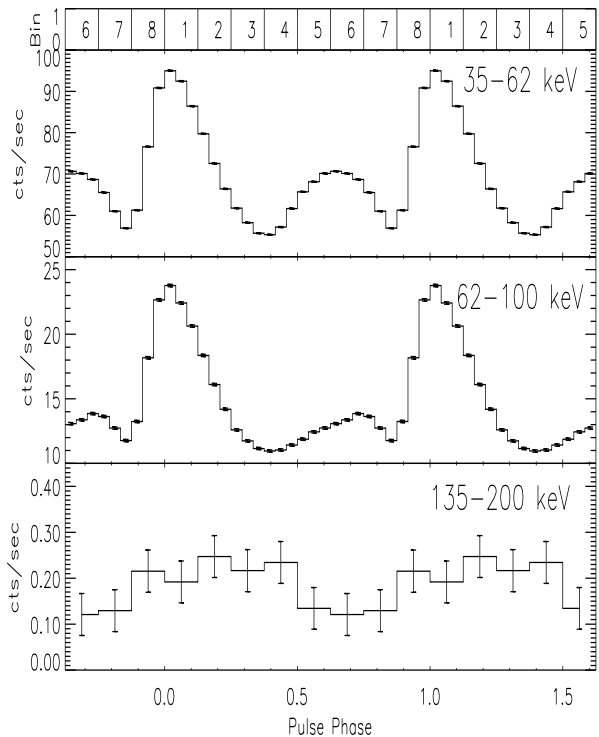,width=0.49\textwidth,height=11cm}
  \caption[]{Background-subtracted pulse profiles of A~0535+26 in the 
   energy ranges 35-62, 62-100 and 135-200 keV. The former two have 24 
phase bins, the latter has been rebinned to 8 phase bins. Note suppressed 
zero in upper two panels. Numbering of 
phase bins is displayed on top.}
\end{figure}

\section{Observations and Data Analysis}

OSSE is one of the 4 instruments on CGRO. It
is a phoswich-type collimated scintillation counter
sensitive between 35~keV and 10~MeV. It consists of four identical detectors
which are collimated to a field of view of 3.8 $\times$ 11.4 degrees (FWHM). 
The total effective area at 511~keV is 2000 cm$^2$. The energy resolution 
is 23\% (FWHM) at 50 keV and 15\% at 100 keV.
Observations are performed by rocking the four detector modules
between source and source-free background
fields on either side of the source field along the scan axis in
a predefined sequence. A detailed description of the instrument
and the data analysis techniques is given in Johnson et al. (1993). 
Following Grove et al. (1995), we include a 3\% systematic error in 
spectral measurements.

The onset of a giant outburst of A~0535+26 was detected by BATSE in January 
1994 (Finger et al. 1994). An observation with OSSE was performed from 
1994 Feb 8-17. Phase averaged spectral analysis with the detection of the 
cyclotron line has been presented by Grove et al. (1995). 
The pulse profiles in the OSSE energy range have already been discussed by 
Maisack et al. (1996): the profile has two peaks, the so-called "main 
pulse" 
centered on phase 0 is more prominent and has a harder spectrum, whereas 
the secondary pulse at phase 0.6 has a softer spectrum (see also 
Kretschmar et al. 1996). The pulse profiles in the energy ranges 35-62, 
62-100 and 135-200 keV are shown in Fig.1. The shape of the profile is 
similar in all three energy bands. 

\section{Spectral fits}

For our phase resolved spectral fits, we use 8 phase bins (corresponding to
$\approx$ 13 sec/bin). This number was chosen to ensure that there is a 
statistically significant signal above 100~keV in each phase bin. 
For the spectral fits, we use the same model as Grove et 
al. (1995) for the phase averaged spectrum, given by Tanaka (1986), in 
which the continuum is described by an exponentially truncated power law 
modified by two Lorentzian absorption lines: 

\begin{equation}
 N(E) = I_{70~keV} (E/70)^{-\alpha} \exp(-E/E_F) \exp(-B)
\end{equation}

where
\begin{equation}
   B =  \tau_1 \frac{W^2 (E/E_{C})^2}{(E-E_{C})^2 + W^2} 
       + \tau_2 \frac{(2W)^2 (E/2 E_{C})^2}{(E-2 E_{C})^2 + (2W)^2}
\end{equation}

Here, $I_{70~keV}$ is the continuum normalisation at 70~keV, 
$\alpha$ is the photon index and 
$E_F$ is the e-folding energy of the continuum, $E_{C}$ the line centroid 
energy, $W$ and $2\times W$ the half widths (HWHM) 
and $\tau_n$ the optical depths of the 
individual cyclotron lines. The energies are given in keV. 
The choice of Lorentzian lines is motivated by the need for an 
analytic formula. While the profiles are expected to have more complicated 
shapes (e.g. Araya and Harding 1996), the energy resolution of OSSE at 
100 keV is 15\%, crude enough to justify the use of such a 
simple approach.

\begin{table*}[t]\centering
\begin{tabular}[c]{cccccc}
\hline
Phase Bin & $\chi_{\nu}$ & $I_{70 keV}$ & $E_F [keV]$ & $E_C [keV]$ & $\tau_2$ \\
\hline
\hline
 1 & 0.80 & 3.69$\pm$ 0.07 &  28.5$\pm$   0.8 & 114.9$\pm$   0.7 &  2.55$\pm$  0.11\\
 2 & 0.75 & 3.42$\pm$ 0.06 &  27.2$\pm$   0.7 & 115.6$\pm$   0.7 &  2.31$\pm$  0.10\\
 3 & 0.72 & 2.30$\pm$ 0.04 &  23.5$\pm$   0.6 & 116.8$\pm$   1.3 &  1.78$\pm$  0.11\\
 4 & 0.92 & 1.94$\pm$ 0.05 &  23.3$\pm$   0.8 & 111.6$\pm$   1.3 &  1.66$\pm$  0.12\\
 5 & 0.55 & 2.05$\pm$ 0.05 &  22.8$\pm$   0.6 & 109.9$\pm$   1.0 &  1.74$\pm$  0.09\\
 6 & 0.68 & 2.24$\pm$ 0.06 &  22.6$\pm$   0.6 & 107.8$\pm$   0.9 &  1.63$\pm$  0.10\\
 7 & 0.69 & 2.34$\pm$ 0.05 &  23.6$\pm$   0.7 & 110.8$\pm$   0.9 &  1.85$\pm$  0.10\\
 8 & 0.95 & 2.62$\pm$ 0.06 &  25.5$\pm$   0.8 & 115.2$\pm$   1.2 &  2.37$\pm$  0.15\\
\hline
\end{tabular}
\caption{Best fit parameters for the phase resolved data in 8 phase bins.  
$I_{70 keV}$ is the continuum intensity at 70~keV in $10^{-3}$~ photons 
(cm$^2$ s keV)$^{-1}$, $E_F$ the e-folding energy in keV, $E_C$ the line 
centroid energy in keV and $\tau_2$ the depth of the line.}
\end{table*}

\begin{figure}[t]
  \psfig{figure=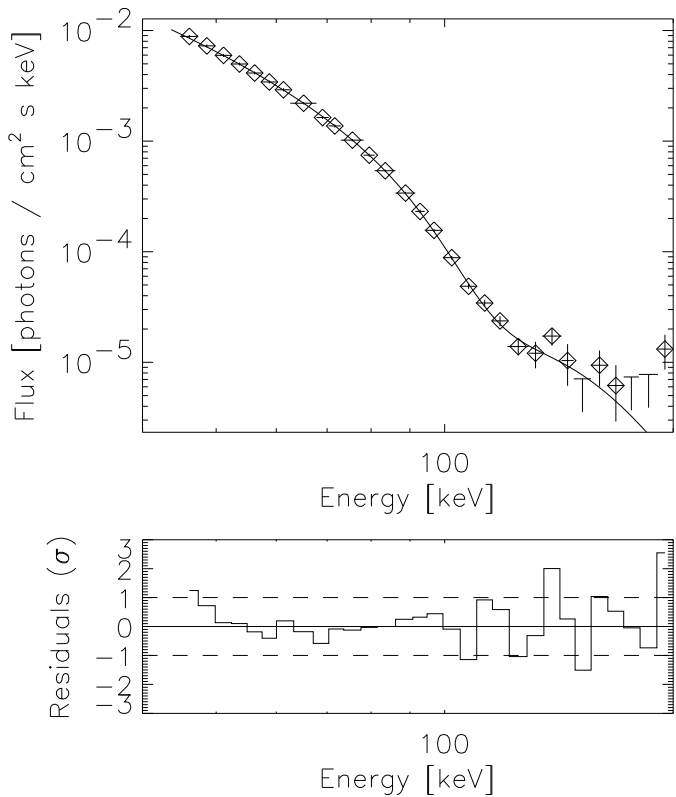}
  \caption[]{Best fit photon spectrum and fit residuals for phase bin 1}
\end{figure}

Analysis of the pulse shape and hardness ratios has shown that the 
spectrum varies with phase (Maisack et al. 1996). 
To verify this in our detailed phase resolved spectral analysis, we first 
attempted to fit the best-fit continuum of the phase averaged spectrum from 
Grove et al. (1995) with the normalisation as the only free parameter to 
each individual phase spectrum. This yields unacceptable fits (probability 
$<$ 1\%) for most phase bins, confirming that the spectrum changes 
with phase. 

Next, we tried to fit a simple continuum without line structures to the 
individual phase spectra, i.e. the only free parameters in this model are the 
normalisation, power law index and e-folding energy. This model yields 
marginally acceptable fits (probability $<$ 10\%) only for the two phase 
bins with the lowest intensity. If the power law index is fixed at the 
TMM/HEXE value of 1.2 (see below for an extended justification of this 
choice), no acceptable fits are achieved. This is convincing evidence that 
additional spectral features are present at every phase. 

Next, we tried to leave all parameters free for 
each phase spectrum, i.e. to allow for two line features. 
The model yields acceptable fits for all phase spectra; 
however, the statistical quality of individual phase spectra is not 
sufficient to derive meaningful constraints on the parameters. 
As in the phase averaged spectrum, the depth of the first line is 
consistent with zero, i.e. the 50~keV feature seen by HEXE (Kendziorra et 
al. 1994) is not required at any phase by the OSSE data, but serves only to 
flatten the continuum at low energies. 
In these fits with all parameters free, the continuum 
parameters in particular (i.e. power law index and 
e-folding energy) cannot be constrained 
since the OSSE energy range $>$~35~keV provides no good handle on the power 
law part of the spectrum. Since both phase averaged and phase resolved 
TTM/HEXE spectra (Kretschmar et al. 1996) are consistent with $\alpha$=1.2, 
with no evidence for variability with phase and overall intensity, we adopt 
this value as a fixed parameter for the OSSE spectral fits. 
This means that the e-folding energy 
carries the information about the hardness of the continuum. 
Moreover, the energy resolution of the OSSE instrument (see Johnson et al. 
1993) is not sufficient to independently constrain the depth and HWHM of 
the line. We therefore keep the HWHM of the line at $\approx$~50~keV 
fixed at 15~keV (this value is the average HWHM from fits of all phase bins, 
and consistent with the spectrum of each individual phase bin) 
and leave the line centroid energy and the depth as the 
free parameters. Note that the HWHM of the line at $\approx$~110~keV is 
then 30~keV.

We therefore fit all individual spectra with line 
width and power law index fixed, and the e-folding energy, 
line centroid, line depth and continuum normalisation 
free. This yields acceptable fits for all phase spectra.
Again, a CRF at $\approx$ 50~keV is possible, but not required, and the 
2$\sigma$ upper limit for the depth is $\tau_1$=0.15.

Since  all phase bin fits are consistent with no CRF at 50~keV, we keep the 
depth of the 50 keV feature fixed at zero and only fit a CRF at $\approx$ 
110~keV (with HWHM 30~keV). We find that such a CRF is then required at 
every phase, and that 
the line centroid energy and line depth vary. 
The fit results and parameters are listed in Table 1 and displayed in Fig.3. 
As an example, the unfolded spectrum and fit residuals for bin 1 are shown 
in Fig. 2.

\section{Variation of parameters with phase}

In this section, we discuss the dependence of the individual fit parameters 
on phase for the fits with a CRF at $\approx$~110~keV, the photon index and 
line width fixed and no CRF at $\approx$~50~keV. 
We find that the continuum is hardest during the main pulse, with an 
e-folding energy of up to 28~keV, while the secondary pulse and 
interpulse phases only have an e-folding energy of $\approx$ 23~keV. 
This parameter traces the continuum intensity at 70~keV well. 
The line centroid energy is also 
found to vary with phase. During the main pulse, the centroid energy 
is 116 keV, compared to 109~keV during the secondary pulse.

\begin{figure}[t]
  \psfig{figure=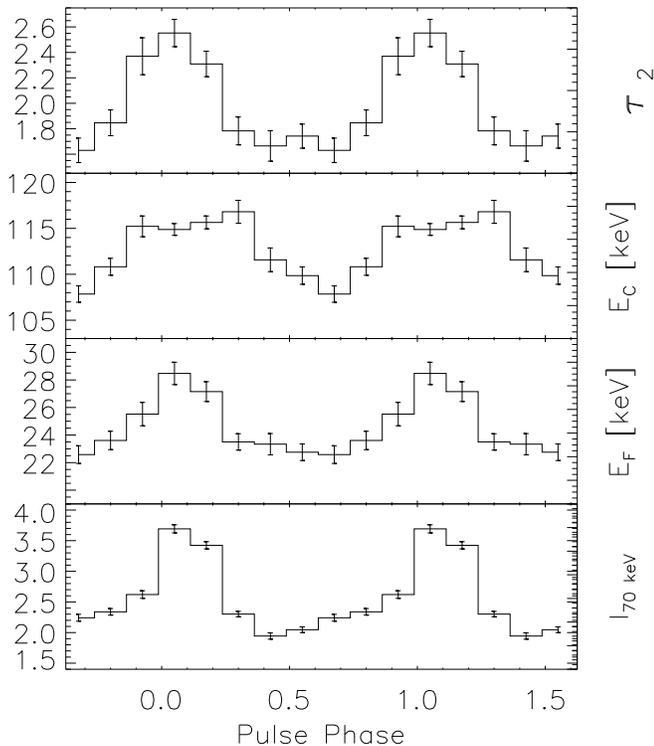}
  \caption[]{Dependence of fit parameters on phase.}
\end{figure}

From Fig.3, it can be seen that continuum 
intensity $I_{70~keV}$, the e-folding energy $E_F$ and the line depth 
$\tau_2$ have essentially identical phase dependence.
The line centroid energy, $E_C$, also varies with phase, being highest during 
the main peak, however, the behaviour of this parameter with phase does 
not exactly 
trace that of the other parameters. The line centroid energy is 
already at its high level of 116 keV before the onset of the 
main pulse, and remains at that level for about half the period. Using 16 
phase bins to achieve better temporal resolution across the phase, one 
finds that the increase in the line centroid energy is indeed rather abrupt, 
with a more gradual decline after the main pulse. 

\subsection{Variability during the observation}

As has already been shown for the phase averaged spectrum by Maisack et al. 
(1996), the spectrum becomes softer as the intensity increases during the 
10 days of the OSSE observation. To show that this is the case for every 
individual phase bin, we divide the observation into two halves of 5 days 
each (Feb 8-12 and Feb 13-17). As can be seen from 
Fig. 4, the e-folding energy is lower (i.e. the 
spectrum is steeper) in the later, brighter half of the observation. 

It is a useful exercise to check the variations of the line centroid energy 
between the early and late halves of the observation. Changes in that 
parameter cannot be due to a sudden change in the neutron star's magnetic 
field and instead convey the systematic uncertainties 
introduced by the varying continuum spectra. The differences between the best 
fit line centroid energies are shown in Fig. 5 
in the same fashion as those for the 
e-folding energy in the previous figure. One can see that the 
phase-dependent trends persist, and that the absolute value of the centroid 
energy 
is subject to a systematic uncertainty of less than 2~keV depending on the 
steepness of the continuum. 

\section{Discussion}

As in the phase averaged analysis, we find a cyclotron feature 
at $\approx$ 110~keV in each of 
the 8 phase bins used for this analysis, whereas a 
line at half this energy is possible but never required. Nagel (1981) 
predicts that the pulse profile changes from double peaked below the 
cyclotron energy to single peaked above for pencil beam configuration at 
moderate optical depth (for 
which we will argue below). 
As there is no sign of a change 
in beam pattern at the putative line energy of $\approx$50~keV, 
we conclude that the feature at 110 keV is the fundamental line. The OSSE 
data above 110~keV, i.e. in the energy range 135-200~keV (see Fig. 1), however, 
are not statistically significant enough to determine 
the possible presence of the secondary peak.

\begin{figure}[t]
  \psfig{figure=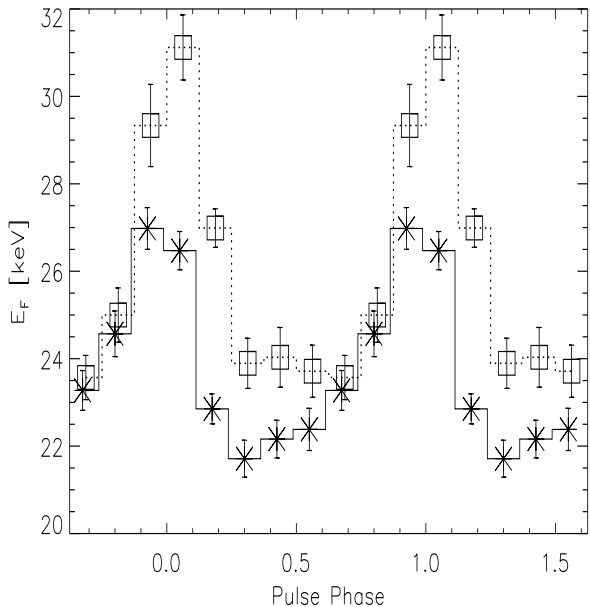,width=0.49\textwidth,height=9cm}
  \caption[]{Comparison of best fit e-folding energies for early (squares) 
and late (asterisks) parts of the observation.}
\end{figure}

\begin{figure}[t]
  \psfig{figure=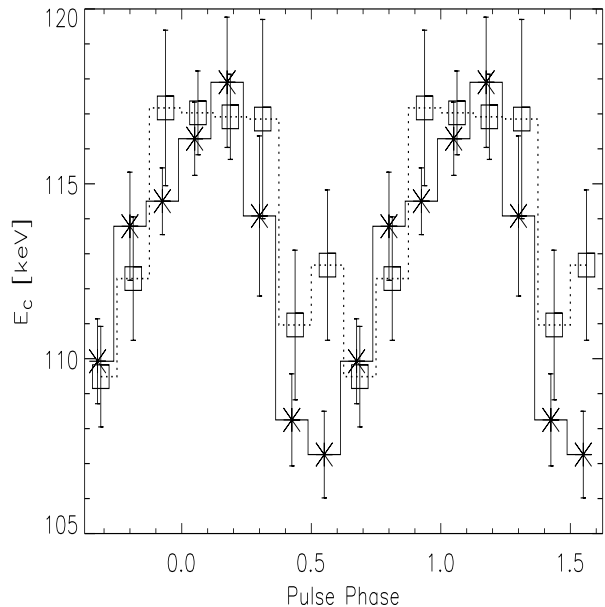,width=0.49\textwidth,height=9cm}
  \caption[]{Comparison of best fit line centroid energies for early 
(squares) and late (asterisks) parts of the observation.}
\end{figure}

The interdependence of the individual spectral parameters can be used to obtain 
information about the emission mechanisms and emission patterns of 
A~0535+26. We find that the line centroid energy is positively correlated 
with the intensity. This is generally expected from a pencil beam emission 
pattern (e.g. Nagel 1981, Meszaros and Nagel 1985). Supporting evidence for 
this comes from the positive correlation between continuum intensity and 
spectral hardness, which is also an indication of a pencil beam 
(Nagel 1981). Additional constraints can be derived from the line 
parameters: as photons propagating parallel to the field lines are most 
easily scattered, the line deficit is expected to be highest as one views 
directly onto the magnetic poles. As this line deficit (parameterised by 
the line depth) varies in phase with the line centroid energy, this is yet 
another indicator for the pencil beam pattern. 

The line centroid energy is highest 
when the emission region is 
viewed parallel to the magnetic field, and is expected to vary with angle 
$\theta$ (Harding and Daugherty 1991) as

\begin{equation}
 h\nu = \frac{m_e c^2}{sin^2 \theta} (\sqrt{1+2 n B sin^2 \theta} -1)
\end{equation}

For a line at $\approx$ 110~keV, the expected variation with angle is thus 
10~keV, approximately the range that is observed. 
The width of the line is 
expected to vary due to thermal Doppler broadening (Harding 1994) as 

\begin{equation}
 \Delta \omega = \omega \sqrt{(2kT/m_e c^2)} cos(\theta)
\end{equation}

where kT is the electron temperature. An electron temperature of 
$\approx$ 7-10~keV can be 
estimated from the cutoff in the X-ray spectrum, which occurs betwen 20 and 
30~keV (Kretschmar et al. 1996). According to the 
above equation, the line at $\theta$=0 would then be broadened by 20~keV. 
Since the HWHM of the 110~keV line is fixed at 30~keV, this is consistent 
with our observations.

A full account of the effects of gravitational light bending is also 
required to fully model the emission pattern (e.g. Meszaros and Riffert 
1988, Riffert and Meszaros 1988) and to determine whether the 
second peak is due to emission from the other pole or not. Modelling of 
these effects has been performed by Kraus et al. (1995) for Cen~X-3.  
However, unique solutions of this model can only be derived if there exist 
profiles at different energies that display 
significantly different pulse shapes. 
The profiles obtained between 35 and 200 keV in this observation are too 
similar for this type of analysis, but future observations by SAX or XTE 
could be used to address this issue.

Our finding that the emission pattern is most likely a pencil beam is 
surprising in view of the high luminosity of A~0535+26 
during the outburst, which 
exceeds 10$^{36}$ ergs/s even above 45 keV (Grove et al. 1995). For 
X-ray luminosities in excess of several $10^{36}$ ergs/s, one generally 
expects a radiative shock which is optically thick in the field 
direction, resulting in a fan beam emission pattern from an accretion 
cylinder (also called accretion column), whereas for lower luminosities a 
collisionless shock results in an optically thin slab which emits a 
pencil beam pattern (e.g. Basko and Sunyaev 1976). 
We point out that contrary to these expectations, Her~X-1 also shows 
these signatures of pencil beam emission (e.g. Harding 1994). The high optical 
depth parallel to the field lines expected for the high luminosity case 
could be circumvented by {\it photon bubbles}, which can 
rise and escape from the polar cap through the accretion 
column (Klein et al. 1996a). Such 
bubbles have been used to  explain QPO seen in GRO~J1744-28 and Sco X-1 
(Klein et al. 1996b). We point out that similar QPO have been detected from 
A~0535+26 by BATSE during the 1994 giant outburst (Finger et al. 1996).

\subsection{About the possible 50~keV feature}

The CRF at $\approx$ 50~keV which was found in the TTM/HEXE data -- albeit 
at low significance -- of the 
1989 giant outburst is not required by the OSSE data, but serves to flatten 
the continuum at low energies. The lack of a change of the pulse profile 
from double to single peaked above this energy (Nagel 1981) argues 
against a CRF at this energy (under the assumption of a pencil beam 
emission pattern). Analysis of the line 
shape in the OSSE data by Araya and Harding (1996) also provided evidence 
that the CRF at 110~keV is the fundamental line. Since $kT << h\omega_{B}$, 
most electrons will populate the ground Landau state (Harding 1994), and 
the fundamental line should be substantially deeper than a harmonic. This 
would require that a possible fundamental at 50~keV would have to be 
favourably filled by photons generated by photon spawning, or hidden by a 
complicated geometry (Araya and Harding 1996).

\section{Conclusions}

We have studied the phase resolved spectrum of A~0535+26 during a giant 
outburst. 
While energy dependent pulse shapes and phase resolved spectra do not 
support the presence of a CRF at $\approx$ 50~keV, it cannot be completely 
ruled out, so this issue needs to be addressed by instruments with better low 
energy coverage, e.g. SAX or XTE. We observe the 
feature at 110~keV at each 
of the 8 phase bins used. The interdependence of individual model parameters 
indicate a pencil beam emission pattern, contrary to expectations from high 
luminosity sources. 
Also taking into account 
gravitational redshifts, we expect the magnetic 
field of A~0535+26 to be in excess of 10$^{13}$ G.

This observation was possible due to the outstanding brightness of A~0535+26 
during a giant outburst. Taking into account the steepness of X-ray binary 
pulsar spectra at high energies, detections of even higher energy CRFs will 
be a difficult task even for future instruments with a sensitivity which 
is expected to be a factor of 10 better than achieved in this observation. 

\acknowledgements

This work was supported under DARA grant 50 OR 92054 and NASA contract 
\mbox{S-10987-C}.

\end{document}